\documentclass{article}
\usepackage[x11names, svgnames, rgb]{xcolor}

\usepackage[english]{babel}

\usepackage[letterpaper,top=2cm,bottom=2cm,left=3cm,right=3cm,marginparwidth=1.75cm]{geometry}

\usepackage{amsmath}
\usepackage{amssymb}
\usepackage{graphicx}
\usepackage{booktabs}
\usepackage{url}
\usepackage[colorlinks=true, allcolors=blue]{hyperref}

\title{KnowTeX: Visualizing Mathematical Dependencies}

\author{Elif Uskuplu\\
Indiana University, Bloomington, IN 47405, USA\\
\texttt{euskuplu@iu.edu}
\and
Lawrence S.~Moss\\
Indiana University, Bloomington, IN 47405, USA\\
\texttt{lmoss@iu.edu}
\and
Valeria de Paiva\\
Topos Institute, Berkeley, CA 94704, USA\\
\texttt{valeria.depaiva@gmail.com}}

\begin{document}
\maketitle

\begin{abstract}
Dependency graphs that show how definitions, theorems, and proofs relate to each other are valuable for understanding the structure of mathematical texts. Existing tools such as Lean Blueprint and plasTeXdepgraph generate such graphs within formal proof ecosystems, but they require familiarity with proof assistants or specific compilation pipelines. We present KnowTeX, a standalone Python tool that extracts dependency graphs directly from \LaTeX\ sources without requiring any external framework. KnowTeX supports two complementary modes: a manual mode where authors annotate their source with lightweight commands compatible with Lean Blueprint, and an infer mode that automatically discovers dependencies through a layered system of deterministic and heuristic rules. The tool handles multi-file projects, detects cycles, applies transitive reduction, and exports graphs in DOT, TikZ, and PNG formats with an interactive preview. We evaluate KnowTeX on several mathematical texts and discuss how it complements recent tools such as LeanArchitect, which operates from the Lean side, while KnowTeX works entirely on the \LaTeX\ side without requiring any formalization.
\end{abstract}


%% ============================================================
\section{Introduction}
%% ============================================================

Mathematical writing is inherently complex. Definitions, theorems, and proofs form dense networks of interdependent ideas, where results often build on earlier ones in subtle and intertwined ways. In textbooks and lecture notes, these connections are spread across multiple chapters and sections, making it difficult to perceive the overall structure of the knowledge being presented.

Formal proof assistants such as Lean~\cite{demoura2015}, Agda~\cite{norell2009}, or Rocq\footnote{Formerly known as Coq.}~\cite{bertot2004} make every dependency explicit: each proof must declare precisely which results it uses, since omitting even a single dependency would render the proof invalid. However, this level of detail can be overwhelming for humans, and it requires a full formalization of the mathematical content.

Between the implicit structure of informal texts and the exhaustive precision of formal systems lies a useful middle ground: dependency graphs. A dependency graph represents mathematical statements as nodes and their conceptual or logical relationships as edges, providing a readable overview of how results depend on one another. Such visualizations make the structure of a text more transparent without requiring formalization.

In the first version of this preprint (arXiv:2601.15294v1), we introduced KnowTeX as a tool for generating dependency graphs from \LaTeX\ sources using explicit author annotations. Since then, KnowTeX has been substantially extended. The current version supports two complementary modes of operation: a \emph{manual mode} using explicit annotation commands compatible with Lean Blueprint, and a new \emph{infer mode} that automatically discovers dependencies through a layered system of deterministic and heuristic rules. In addition, the tool provides an interactive GUI for reviewing and editing inferred edges, handles multi-file \LaTeX\ projects, detects dependency cycles, and applies transitive reduction for cleaner graph output.

Our contributions in this paper are as follows:
(1)~We describe the design of KnowTeX's two-mode architecture: manual annotation and automatic inference with eight layered rules (D1--D4, H1--H4).
(2)~We present quantitative evaluation on four case studies: two textbook-scale documents with manually annotated gold standards, a corpus-scale experiment on 611 definitions from the MathGloss dataset~\cite{mathgloss}, and a validation against an existing Lean Blueprint project.
(3)~We compare KnowTeX with related systems, including Lean Blueprint, plasTeXdepgraph, LeanArchitect, and Trouver, and discuss how KnowTeX occupies a complementary position.

KnowTeX is publicly available as open-source software.\footnote{\url{https://github.com/ElifUskuplu/KnowTex}}

%% ============================================================
\section{Background: The Role of Dependency Graphs}
\label{sec:background}
%% ============================================================

Dependency graphs provide practical value at multiple stages of mathematical work. For the end reader, whether a student or a researcher approaching an unfamiliar text, a dependency graph serves as a study map: it shows which definitions and lemmas are needed for a target result, making it possible to follow a selective path through the material rather than reading every page from the beginning.

For the author, generating a dependency graph from a draft can act as a structural review tool before the final version. An isolated node that is not the result of an inference error signals that a definition or lemma has no connection to the rest of the text, prompting the author to either integrate it more explicitly or remove it. A cycle in the graph, when it does not stem from a tool error, may reveal a genuine circular dependency between two results, which is a logical issue worth resolving before publication. More broadly, the graph makes visible which results carry the most weight (many outgoing edges) and which parts of the text are structurally thin (few connections), giving the author concrete feedback on the organization of the document.

The connection to formalization is also direct. In large formalization projects such as PFR, PrimeNumberTheoremAnd, and Carleson (see Sections~\ref{sec:related} and~\ref{subsec:eval-pfr}), the Lean Blueprint dependency graph is the main tool for tracking what has been formalized and what remains. A dependency graph extracted from a \LaTeX\ source is, in rough terms, a formalization road map. A researcher planning to formalize a result in Lean or another proof assistant can read off the minimal set of prerequisites from the graph, state them with \texttt{sorry} placeholders, and focus effort on the main theorem. If the graph shows that the main result does not actually depend on certain lemmas, the author may discover that the informal text contains unnecessary material, leading to a cleaner paper as well.

Finally, dependency graphs can serve as structured metadata for mathematical corpora. The need for such datasets has been discussed by several researchers~\cite{collard2024,depaiva2023,fong2021,welleck2021}. When combined with external identifiers such as Wikidata QIDs, the resulting graph becomes a knowledge resource that links informal mathematical text to the broader semantic web. We explore this direction in Section~\ref{subsec:eval-corpus}, where we apply KnowTeX to a corpus of 611 definitions with Wikidata alignment, producing dependency metadata that could support tasks such as prerequisite prediction, concept clustering, or automated curriculum planning.

%% ============================================================
\section{Related Work}
\label{sec:related}
%% ============================================================

Several tools have explored ways to generate dependency graphs from mathematical documents.

\paragraph{Earlier work on structure in \LaTeX\ documents.}
Extracting structural information from \LaTeX\ sources is not new. Kamareddine et al.~\cite{kamareddine2007} studied the narrative structure of mathematical texts, including the dependencies between their parts, within the MathLang framework. The sTeX ecosystem of Kohlhase and collaborators\footnote{See, e.g., \url{https://ceur-ws.org/Vol-3377/natfom4.pdf}.} extends \LaTeX\ toward semantic markup, and its explicit module imports make dependency structure available to downstream tools. sTeX asks authors to adopt new writing conventions. KnowTeX takes the opposite stance: it does not change \LaTeX, it works on the author's existing source, and it respects their own macros, so a mathematician does not need to learn a new system to use it. Our aim is not to change how mathematicians write, but to use what they already write and give them a way to check how well structured it is. The same reasoning is why KnowTeX does not build on plasTeX, discussed next.

%% TODO (promised in the CICM rebuttal, citations to be verified before adding):
%% - Isabelle theory/session dependency graphs (e.g. https://isabelle.in.tum.de/library/HOL/HOL-Analysis/session_graph.pdf)
%% - Fabian Huch's work on classifying dependencies in the Archive of Formal Proofs
%% - zbMATH Open indexing work (Christian Baer et al., https://www.newton.ac.uk/seminar/50258/)

\paragraph{Lean Blueprint and plasTeXdepgraph.}
The most influential system for representing mathematical dependencies is Massot's Lean Blueprint\footnote{\url{https://github.com/PatrickMassot/leanblueprint}}, which links informal mathematical text with corresponding formal proofs in the Lean theorem prover. A simplified variant, plasTeXdepgraph\footnote{\url{https://github.com/PatrickMassot/plastexdepgraph}}, is a plugin for plasTeX~\cite{plastex} that generates dependency graphs without requiring the Lean prover.

PlasTeX processes \LaTeX\ documents into an XML-DOM-like object, and the plasTeXdepgraph plugin compiles this into an HTML format with an interactive dependency graph. This interactivity is effective when the \LaTeX\ source is written for plasTeX, but compiling legacy or complex documents with custom macros can be challenging. In our experiments with Leinster's \textit{Basic Category Theory}~\cite{leinster2014}, successful compilation required extensive editing and removal of non-essential macros.
The Lean Blueprint system, on the other hand, extends this HTML-based foundation by connecting dependency graphs with formal proof development, using visual cues to indicate proof status. These features are powerful but Lean-centered and presuppose an existing formalization.

\paragraph{LeanArchitect.}
A recent development is LeanArchitect~\cite{leanarchitect2025}, which automates the synchronization between Lean~4 code and \LaTeX\ blueprint documents. Using a \texttt{@blueprint} attribute, developers tag Lean declarations with \LaTeX\ labels and natural-language descriptions. The system automatically infers dependency relations and formalization status (e.g., whether a proof still contains \texttt{sorry}) by traversing the Lean code, and exports this information as \LaTeX\ fragments. The actual dependency graph is then produced downstream by leanblueprint from these generated \LaTeX\ artifacts.

LeanArchitect addresses a real pain point in large formalization projects: keeping documentation synchronized with evolving code. It has been applied to projects such as \textit{PrimeNumberTheoremAnd}\footnote{\url{https://github.com/AlexKontorovich/PrimeNumberTheoremAnd}} and \textit{Carleson}\footnote{\url{https://github.com/fpvandoorn/carleson}}. However, it is tightly integrated with the Lean~4 ecosystem and requires an active formalization project. KnowTeX, by contrast, works purely at the \LaTeX\ level and requires no proof assistant, making it suitable for authors who want dependency visualization without formal verification.

\paragraph{Trouver.}
The Trouver system\footnote{\url{https://github.com/hyunjongkimmath/trouver}} takes a machine-learning approach to extracting dependency information from \LaTeX\ sources. Unlike KnowTeX, which relies on explicit commands (manual mode) or structural rules (infer mode), Trouver infers relationships automatically from textual and linguistic patterns. This automation makes it attractive for large-scale or legacy corpora, but inferred links may miss subtle logical dependencies or introduce spurious ones. KnowTeX prioritizes explicit author control and reproducibility over full automation, providing a complementary rather than competing model.

\paragraph{ScienceStack.}
ScienceStack\footnote{\url{http://www.sciencestack.ai}} is a commercial service that provides HTML rendering and dependency graphs for \LaTeX\ papers. However, it uses closed-source software and requires payment, limiting accessibility and reproducibility.

%% ============================================================
\section{KnowTeX: Design and Architecture}
\label{sec:design}
%% ============================================================

KnowTeX is a standalone Python application with a GUI built using Tkinter. It takes \LaTeX\ projects as input and produces dependency graphs in DOT (Graphviz), TikZ, and PNG formats, with an interactive preview window. The processing pipeline has five stages: file expansion, structure detection, parsing, edge extraction, and graph construction.

\subsection{File Expansion and Structure Detection}
\label{subsec:expansion}

KnowTeX recursively expands \verb|\input|, \verb|\include|, \verb|\import|, \verb|\subimport|, and \verb|\subfile| directives to produce a single expanded document. The expansion step handles \verb|\includeonly| filtering, cycle prevention (each file path is visited at most once), and path traversal safety. Files outside the project directory are blocked.
After expansion, KnowTeX detects the document class (\texttt{book} vs.\ \texttt{article}) and identifies chapter or section boundaries. Users can select which chapters or sections to include, enabling localized graph generation.

\subsection{Parsing}
\label{subsec:parsing}

The parser uses pylatexenc to walk the expanded \LaTeX\ AST and extract two kinds of objects: \emph{nodes} (statement-like environments such as Definition, Theorem, Lemma, Proposition, Corollary, Construction, Example, and Remark) and \emph{proofs} (proof environments). KnowTeX does not rely on a fixed list of environment names: every environment found in the source is treated as a candidate statement node, except proof environments and a set of well-known structural ones (figure, table, equation, align, tikzpicture, etc.), which are skipped. Before the graph is built, the user can include or exclude each detected environment and mark which ones are definition-like. A custom theorem-like environment, or a conjecture environment, is therefore detected automatically, whatever its name. Standard reference and emphasis commands are recognized; custom wrapper macros (for example, a user-defined command used in place of \verb|\ref|) are not expanded, but any dependency they hide can still be added in the review step (Section~\ref{subsec:review}).

Each node is assigned a unique label through a resolution hierarchy: (1) explicit \verb|\label{}|, (2) a label derived from emphasized terms (\verb|\emph{}|, \verb|\textit{}|, \verb|\textbf{}|) within the environment, (3) a label derived from \verb|\index{}| entries, or (4) a sequential numeric fallback. Each proof is initially associated with its nearest preceding statement (rule H1) and may be redirected by an explicit proof header such as \verb|\begin{proof}[Proof of Theorem \ref{...}]| (rule D3) or by a \verb|\proves{}| command.

To classify detected environments by kind (definition, theorem, lemma, corollary, and so on), which the heuristic rules and the node styling rely on, KnowTeX supports multiple naming conventions through regular expressions, for example, recognizing shortened forms such as \texttt{thm}, \texttt{def}, \texttt{cor}, or \texttt{lem}.

\subsection{Manual Mode}
\label{subsec:manual}

In manual mode, authors annotate their \LaTeX\ source with two commands:
\begin{itemize}
\item \verb|\uses{label1, label2, ...}| declares that the current statement or proof depends on the listed labels.
\item \verb|\proves{label}| declares that the current proof establishes a particular labeled statement.
\end{itemize}

When \verb|\uses{}| appears inside a statement, the resulting edge is drawn as a \emph{dashed} arrow (conceptual dependency). When it appears inside a proof, the edge is drawn as a \emph{solid} arrow (logical dependency from the proof). This command system follows the same convention used by plasTeXdepgraph and Lean Blueprint, ensuring full compatibility: documents annotated for either system can be processed by KnowTeX without modification.

\subsection{Infer Mode}
\label{subsec:infer}

Infer mode is the main new contribution of KnowTeX compared to the first version of this preprint. It automatically discovers dependencies by applying a layered system of eight rules in a fixed order. Each rule adds edges to the graph, and a shared deduplication set prevents the same edge from being created by multiple rules.

\subsubsection*{Deterministic Rules (D1--D4).}
These rules extract edges from concrete syntactic evidence in the \LaTeX\ source:

\paragraph{D1: Proof cross-references.} For each proof, scan for \verb|\ref|, \verb|\Cref|, \verb|\cref|, and \verb|\eqref| commands. Each referenced label that resolves to a known node produces an edge from that node to the proof's parent statement.

\paragraph{D2: Statement cross-references.} For each theorem-like statement, scan for the same reference commands. Each referenced label produces an edge from the referenced node to the current statement.

\paragraph{D3: Explicit proof target.} When a proof header contains a reference such as \begin{center}
\verb|\begin{proof}[Proof of Theorem \ref{thm:X}]|,    
\end{center}
the proof is associated with the referenced statement rather than the nearest preceding one.

\paragraph{D4: Defined-term matching.} Terms introduced via \verb|\emph{}|, \verb|\textit{}|, \verb|\textbf{}|, or \verb|\index{}| in definition-like environments are extracted and stemmed using the Snowball English stemmer.
%\footnote{\url{https://snowballstem.org/}}
Subsequent statements whose text contains matching stems receive a dependency edge from the defining node. Multi-word terms require contiguous subsequence matching to reduce false positives.

\subsubsection*{Heuristic Rules (H1--H4).}
These rules handle common structural patterns that cannot be captured by syntactic evidence alone:

\paragraph{H1: Default proof association.} Each proof is associated with the most recently encountered theorem-like statement. This default can be overridden by D3 or \verb|\proves{}|.

\paragraph{H2: Corollary to nearest theorem.} A corollary that contains no cross-references and has no existing incoming edge is linked to the nearest preceding theorem or proposition.

\paragraph{H3: Lemma to next theorem.} A lemma is linked to the next theorem or proposition within a configurable gap (default: 3 nodes), reflecting the common pattern where lemmas serve as stepping stones for a subsequent result.

\paragraph{H4: Index-term matching.} Terms from \verb|\index{}| entries are matched across statements using a longest-match-first strategy with \verb+|see{}+ alias resolution and hierarchical term expansion.

The execution order is: H1 and D3 during parsing, then D1, D2, D4, H2, H3, H4 during inference. This ordering ensures that deterministic evidence takes priority over heuristic guesses.

\subsection{Interactive Edge Review}
\label{subsec:review}

After edge extraction, the GUI presents all edges in a table view showing source, target, type (deterministic/heuristic/manual), location (proof/statement/inferred), and rule. Cycle edges are highlighted in red. Users can delete incorrect edges, add new ones manually, and changes trigger immediate cycle re-detection. This interactive step allows users to correct inference errors before generating the final graph.

\subsection{Graph Construction and Export}
\label{subsec:export}

KnowTeX builds a Graphviz directed graph from the filtered nodes and edges. Optionally, transitive reduction is applied to remove redundant edges for cleaner visualization. Cycles are detected using Tarjan's strongly connected component algorithm and highlighted in red. The graph is exported in three formats: DOT (Graphviz source), TikZ (via dot2tex, for direct inclusion in \LaTeX\ documents), and PNG. A zoomable, pannable preview with clickable nodes is available within the GUI.

%% ============================================================
\section{Evaluation}
\label{sec:evaluation}
%% ============================================================

We evaluate KnowTeX through four case studies. In the first two, a group theory document and three chapters of Leinster's \textit{Basic Category Theory}~\cite{leinster2014}, we compare the inferred output against manually annotated gold-standard graphs constructed using explicit \verb|\uses{}| annotations. In the third, we apply KnowTeX to 611 definitions from the MathGloss corpus, where inter-definition hyperlinks serve as the ground truth. In the fourth, we validate KnowTeX against an existing Lean Blueprint project to test compatibility with the Blueprint system.

It is important to clarify what we mean by gold standard in these experiments. In Case Studies 1 and 2, the manual graph was constructed by the first author based on a close reading of the source text. Dependency annotation is inherently subjective: different annotators may disagree on whether a particular result truly depends on another, especially for conceptual (as opposed to proof-level) dependencies. The manual graphs should therefore be understood not as the unique correct answer, but as one informed human judgment against which the inference is measured. This subjectivity is part of why KnowTeX includes an interactive review step: the inferred graph is a starting point, and the final graph reflects the author's own understanding of the text's structure.

\subsection{Case Study 1: Group Theory}
\label{subsec:eval-groups}

The group theory document contains 42 labeled environments (definitions, theorems, lemmas, propositions, corollaries, examples, and remarks). KnowTeX achieved perfect node detection (100\%); the edge-level results are summarized in Table~\ref{tab:eval-groups}.

\begin{table}[htbp]
\caption{Evaluation results for the group theory document.}\label{tab:eval-groups}
\centering
\footnotesize
\begin{tabular}{@{}lr@{}}
\toprule
Metric & Value \\
\midrule
Nodes (manual / inferred) & 42 / 42 \\
Manual edges & 64 \\
Inferred edges & 68 \\
True positives & 54 \\
False negatives (missed) & 10 \\
False positives (extra) & 14 \\
Precision & 0.794 \\
Recall & 0.844 \\
F1 score & 0.818 \\
\midrule
Path-based precision & 0.971 \\
Path-based recall & 0.844 \\
Truly spurious edges & 2 / 68 \\
\bottomrule
\end{tabular}
\end{table}

The direct edge F1 of 0.818 reflects good overall accuracy. A path-based analysis reveals that out of 14 extra inferred edges, 12 are transitively justified by the manual graph: they represent shortcut edges that make explicit what is already implied through longer paths. Only 2 out of 68 inferred edges are truly spurious (no path exists in the manual graph), yielding a path-based precision of 0.971.

The main source of error is the neighborhood of \texttt{def:homo} (homomorphism): the inference connects this definition directly to distant theorems while missing intermediate structural concepts such as kernel, image, and quotient. This pattern occurs because the intermediate concepts are not explicitly referenced via \verb|\ref{}| in the relevant proofs, and their defined terms are too generic for reliable stemming matches.

\subsection{Case Study 2: Basic Category Theory}
\label{subsec:eval-bct}

\begin{figure}[h!]
\centering
\includegraphics[width=\linewidth]{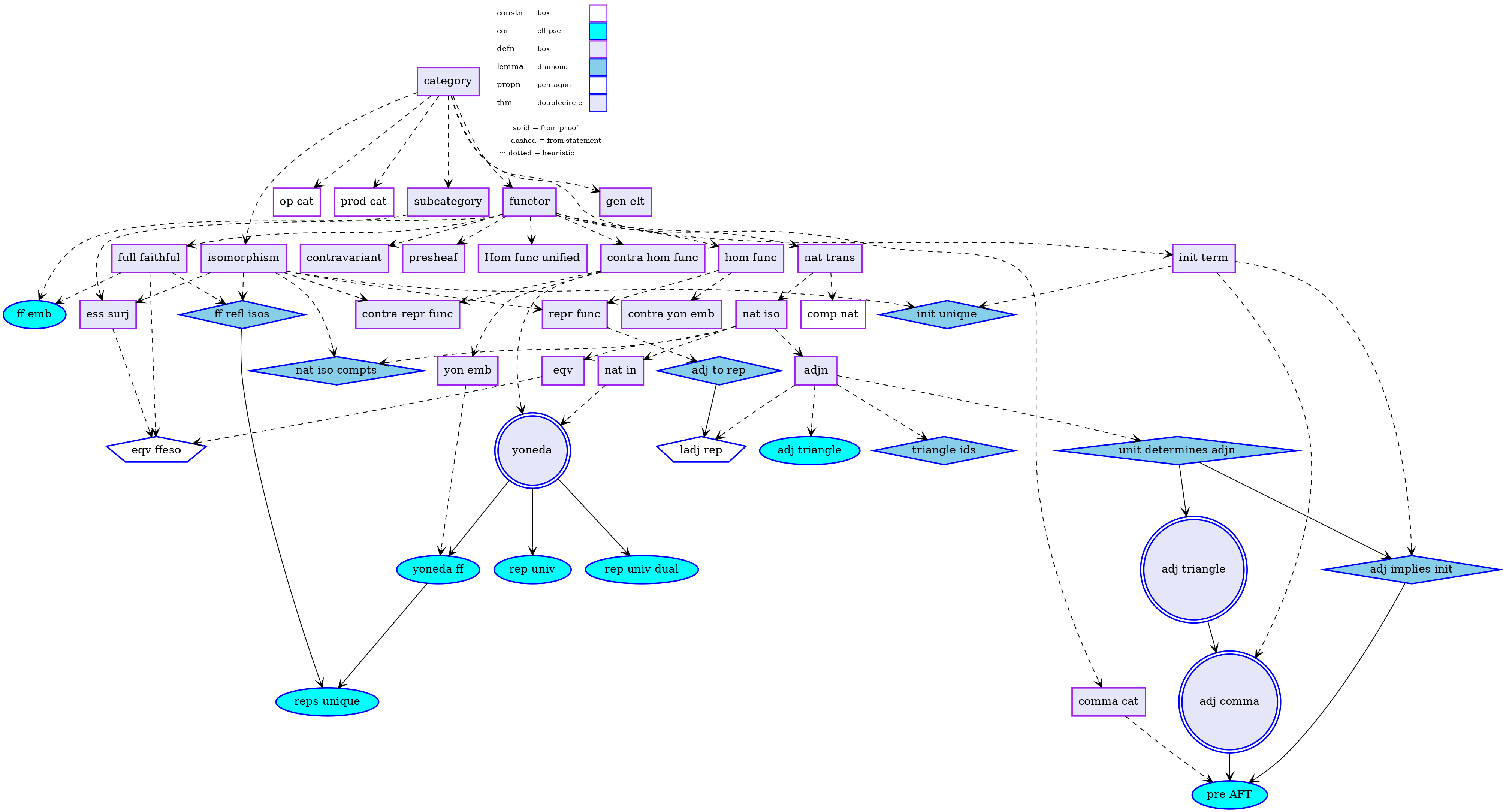}
\caption{Manual (gold-standard) dependency graph for Chapters~1,~2,~and~4 of Leinster's \textit{Basic Category Theory}, constructed using explicit \texttt{\textbackslash uses\{\}} annotations.}
\label{fig:bct-manual}
\end{figure}

\begin{figure}[h!]
\centering
\includegraphics[width=\linewidth]{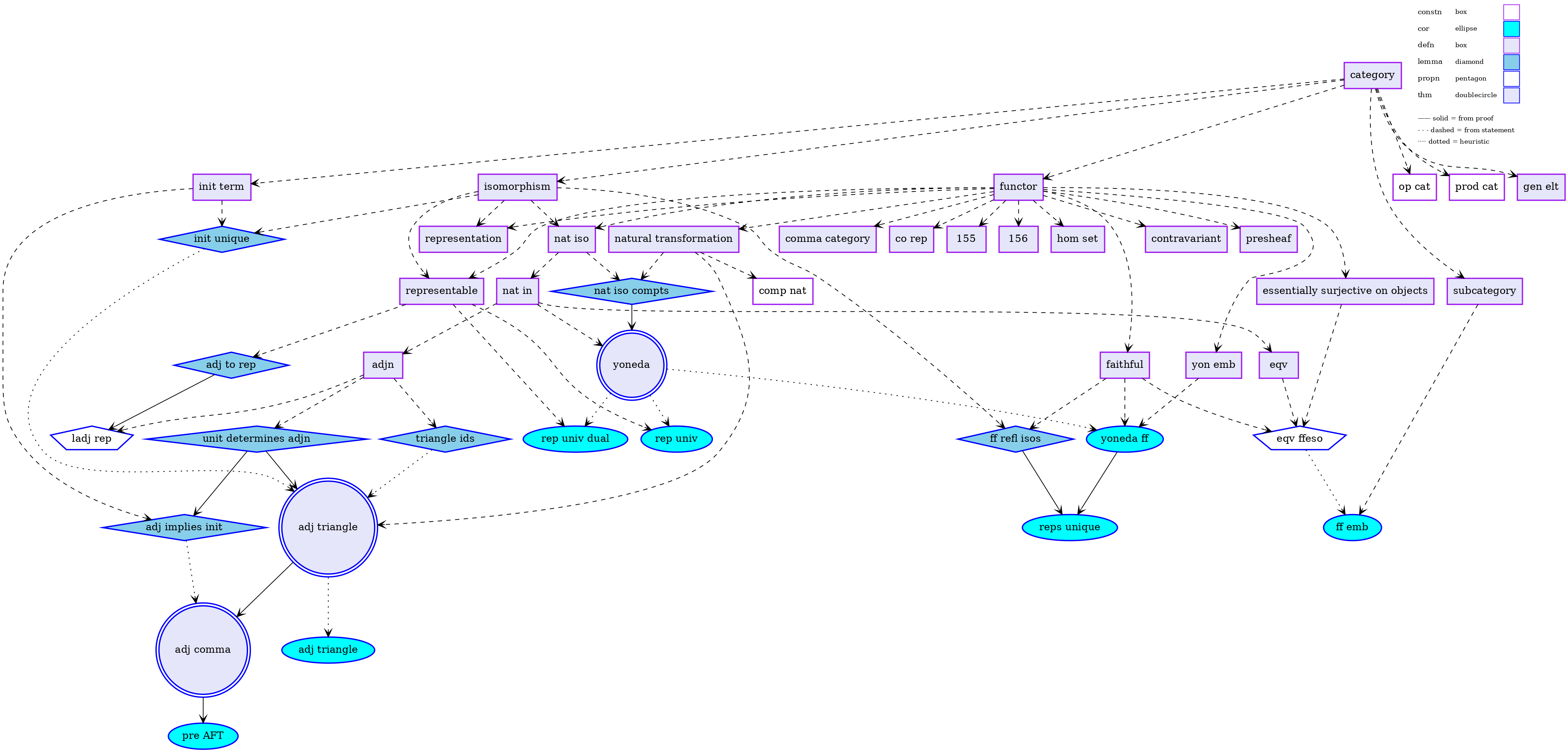}
\caption{Inferred dependency graph for the same chapters, produced automatically by KnowTeX's infer mode without any annotations.}
\label{fig:bct-inferred}
\end{figure}

The three chapters, leading to the adjoint theorems and Yoneda lemma, of Leinster's textbook contain 45 labeled environments. For this case study, we ran infer mode directly on the original \LaTeX\ source provided by the author, without any modifications to the text. The manual gold-standard graph, by contrast, was constructed separately by adding \verb|\uses{}| annotations to a copy of the same source. The full manual and inferred graphs are shown in Figures~\ref{fig:bct-manual} and~\ref{fig:bct-inferred}, respectively. The evaluation is complicated by naming differences: the manual and inferred graphs use different label conventions for 10 semantically identical concepts (e.g., \texttt{defn:nat-trans} vs.\ \texttt{defn:natural-transformation}). We report both raw and normalized results.

\begin{table}[htbp]
\caption{Evaluation results for Basic Category Theory (Chapters~1,~2, and~4).}\label{tab:eval-bct}
\centering
\footnotesize
\begin{tabular}{@{}lrr@{}}
\toprule
Metric & Raw & Normalized \\
\midrule
Matching nodes & 35 / 45 & 45 / 45 \\
Manual edges & 61 & 61 \\
Inferred edges & 65 & 65 \\
True positives & 32 & 46 \\
Precision & 0.492 & 0.708 \\
Recall & 0.525 & 0.754 \\
F1 score & 0.508 & 0.730 \\
\midrule
Reachability precision & -- & 0.752 \\
Reachability recall & -- & 0.880 \\
Edge style accuracy & -- & 93.5\% \\
\bottomrule
\end{tabular}
\end{table}

After normalization, all 45 nodes match, and the edge-level F1 improves from 0.508 to 0.730 (Table~\ref{tab:eval-bct}). The large gap between raw and normalized scores shows that a significant portion of the disagreement comes from naming differences rather than structural errors.

Among the correctly identified edges, 93.5\% also have the correct style classification (solid for proof-level, dashed for statement-level, dotted for heuristic). The main misclassifications involve the Yoneda theorem's corollaries, which are correctly linked but marked as heuristic instead of proof-based. A visual comparison of Figures~\ref{fig:bct-manual} and~\ref{fig:bct-inferred} shows that the overall graph topology is preserved, despite the naming and edge-type differences.

\subsection{Case Study 3: Corpus-Scale Experiment}
\label{subsec:eval-corpus}
To test KnowTeX beyond single documents, we applied it to the Chicago subset of the MathGloss dataset~\cite{mathgloss}, an independently curated glossary that aligns mathematical definitions from multiple sources with Wikidata identifiers. The Chicago subset, drawn from University of Chicago undergraduate mathematics lecture notes, contains 611 definitions in the project's GitHub repository. Each definition is a short markdown text with hyperlinks to other definitions it depends on; 593 of the 611 definitions contain at least one such link. These hyperlinks form a natural ground truth of 1,660 unique dependency edges.

\paragraph{Benchmark design.} Converting this corpus into a format suitable for KnowTeX required careful design decisions that turned out to be critical for meaningful evaluation. The definitions were converted into a single \LaTeX\ file as follows: (1)~inter-definition hyperlinks were preserved as \verb|\ref{}| commands, (2)~each definition's title was inserted as an \verb|\emph{}| term at the beginning of the environment, and (3)~the definitions were topologically sorted so that depended-upon definitions appear earlier in the document, matching KnowTeX's forward-reference guard in the D4 rule. Since the corpus contains only definitions (no theorems, lemmas, or proofs), only rules D2 (statement cross-references) and D4 (defined-term matching) are applicable; the remaining rules do not fire in this setting.

The conversion design matters more than it might seem. In an earlier version of the benchmark, hyperlinks were converted to \verb|\emph{}| text rather than \verb|\ref{}| commands. This had two effects. D2 could not fire at all, since it relies on explicit cross-reference commands; and D4 treated every linked term as a term \emph{defined} by the citing definition, which suppressed the correct edges and produced over two thousand spurious ones, resulting in an F1 of only 0.08. Converting the links to plain text instead, so that only D2 is disabled, gives an F1 of 0.671 for D4 alone (see the ablation row in Table~\ref{tab:eval-corpus}). Restoring the hyperlinks as \verb|\ref{}| commands was not an artificial boost but a correction of a methodological error: the hyperlinks in MathGloss are explicit dependency annotations, and the benchmark should preserve rather than discard them. Similarly, an early version converted all inline bold words to \verb|\emph{}|, which produced single-word terms such as ``algebra'' from a definition titled ``algebra of subsets''. D4's single-word stem matching then fired on every unrelated definition containing that stem, producing hundreds of false positives. Using the definition title as the \verb|\emph{}| term instead ensures that multi-word concepts like ``algebra of subsets'' are matched as phrases, which is much more precise.

We stress that the conversion adds no information: the hyperlinks and the definition titles are already present in the MathGloss markdown files, and we kept them rather than adding anything new. As the results below show, the perfect recall comes entirely from the authors' own hyperlinks through D2, while the titles fed to D4 are what lowers precision. The conversion therefore does not inflate the result; if anything, it lowers it.
\begin{table}[htbp]
\caption{Evaluation results on the MathGloss Chicago corpus (611 definitions).}\label{tab:eval-corpus}
\centering
\footnotesize
\begin{tabular}{@{}lrrr@{}}
\toprule
Rule & Precision & Recall & F1 \\
\midrule
D2 (cross-references) & 1.000 & 1.000 & 1.000 \\
D4 (term matching) & 0.000 & 0.000 & 0.000 \\
All rules combined & 0.793 & 1.000 & 0.885 \\
\midrule
D4 alone (\verb|\ref{}| removed, ablation) & 0.709 & 0.637 & 0.671 \\
\bottomrule
\end{tabular}
\end{table}

\paragraph{Results.} Table~\ref{tab:eval-corpus} shows the per-rule and combined results. The D2 rule recovers all 1,660 ground truth edges with perfect precision and recall. This is expected: the MathGloss hyperlinks were converted to \verb|\ref{}| commands, and D2 extracts exactly these cross-references. The D4 rule adds 433 edges that are not in the ground truth, bringing the combined precision to 0.793 while recall remains at 1.0. The per-rule D4 metrics show zero precision and recall because of the deduplication order: D2 runs first and claims every ground truth edge, so D4 is credited only with edges beyond the ground truth. This is an artifact of the setup, not a measure of D4. To measure D4 on its own, we ran an ablation in which the hyperlinks were converted to plain text (so that D2 has nothing to read) while the text and the ground truth stayed the same. D4 alone then recovers 1,057 of the 1,660 ground truth edges (precision 0.709, recall 0.637, F1 0.671), which is the figure to quote for term matching in the absence of cross-references.

\paragraph{Error analysis.}
We reviewed all 433 edges flagged as false positives through manual inspection and reachability analysis. We read each source and target definition in the MathGloss repository, identifying 220 as genuine dependencies. We then checked whether the target is reachable from the source through a directed path of ground-truth edges. Of the 220 genuine edges, 131 were confirmed reachable. Among the remaining 213 edges initially marked as false positives, 34 proved reachable as well and were reclassified as genuine.

The final classification splits the 433 edges into three groups. First, 165 edges (38.1\%) are genuine dependencies verifiable through ground-truth paths. For example, D4 infers 
\begin{center}
\texttt{def:field}~$\to$~\texttt{def:linear-transformation},
\end{center}
which follows the chain
\begin{center}
    \texttt{def:field}~$\to$~\texttt{def:vector-space}~$\to$~\texttt{def:linear-transformation}
\end{center}
in the ground truth. Second, 89 edges (20.6\%) are genuine dependencies with no ground-truth path, such as
\begin{center}
    \texttt{def:abelian-group}~$\to$~\texttt{def:divisible-group} and \texttt{def:permutation}~$\to$~\texttt{def:alternating-group},
\end{center}
where the target presupposes the source concept but the corpus does not record the link. Third, 179 edges (41.3\%) are true false positives, 170 of them from single-word stem matching. The largest contributors are
\begin{itemize}
    \item \texttt{def:functional} (77 edges, where the stem ``function'' matches unrelated definitions),
    \item \texttt{def:derivation} (20 edges, conflating algebraic derivation with calculus derivative), and
    \item \texttt{def:symmetric} (10 edges, confusing symmetric relation with symmetric group).
\end{itemize}
In total, 254 of the 433 flagged edges (58.7\%) are genuine dependencies rather than errors.

The hyperlink-based gold standard captures only the dependencies that the original authors chose to annotate. A closer look at the 89 genuine edges with no ground-truth path reveals several reasons for their absence: some involve broken hyperlinks in the source data (e.g.\ a link targeting \texttt{abelian} instead of the correct \texttt{abelian\_group}), some refer to the source concept but link to a broader term instead (e.g.\ several definitions mention ``abelian group'' but link only to \texttt{group}, not to \texttt{abelian-group}), and some reflect indirect relationships that the authors may not have considered direct enough to annotate. These 89 edges are therefore not necessarily errors in the ground truth but rather a mix of annotation choices, oversights, and broken references. The benchmark script and per-edge classifications are included in the project repository.

\subsection{Case Study 4: Lean Blueprint Validation}
\label{subsec:eval-pfr}

The previous case studies evaluate KnowTeX against gold standards constructed by the authors. A natural question is whether KnowTeX's manual mode is truly compatible with the Lean Blueprint system it claims to support. To test this, we applied KnowTeX to the Polynomial Freiman--Ruzsa (PFR) conjecture formalization project~\cite{pfr2023}, a large-scale Lean Blueprint project with 218 labeled environments and 403 dependency edges.

\paragraph{Why PFR.}
We chose PFR over other Lean Blueprint projects because its \LaTeX\ source contains all necessary \verb|\uses{}| commands for every dependency. In many other Blueprint projects, dependencies are tracked on the Lean side using macros such as \verb|\inputleannode{}| that reference Lean declarations directly, rather than being written as explicit \verb|\uses{}| commands in the \LaTeX\ source. Since KnowTeX operates purely on \LaTeX, it can only process projects where the dependency information is present in the source text. PFR satisfies this requirement.

Notably, the PFR project authors (\url{https://github.com/teorth/pfr}) are not affiliated with our team; their independent use of \verb|\uses{}| annotations demonstrates that dependency tracking is a practice adopted by the broader formalization community.

\paragraph{Three-way comparison.}
We compared three dependency graphs: (1)~the Lean Blueprint graph extracted from the project's published HTML rendering, (2)~KnowTeX's manual mode output from the same \LaTeX\ source, and (3)~KnowTeX's infer mode output from the same source with all annotations removed. Two lemmas in the PFR source carry two \verb|\label{}| commands each: \texttt{rho\_init} and \texttt{rho\_of\_uniform} on the same lemma, and \texttt{rhominus\_nonneg} and \texttt{rhoMinus\_nonneg} on another. KnowTeX selects the first label while the Blueprint selects the second, producing different node identifiers for the same lemma. After normalizing these two pairs, all 218 nodes match across the three graphs.

\begin{table}[htbp]
\caption{Evaluation results for the PFR Lean Blueprint project (218 nodes, normalized).}\label{tab:eval-pfr}
\centering
\footnotesize
\begin{tabular}{@{}lrrrr@{}}
\toprule
Mode & Precision & Recall & F1 & Style acc. \\
\midrule
Manual & 0.998 & 0.998 & 0.998 & 100\% \\
Infer & 0.898 & 0.764 & 0.826 & 96.4\% \\
Infer (path-based) & 0.921 & 0.814 & 0.864 & -- \\
\bottomrule
\end{tabular}
\end{table}

\paragraph{Results.}
Table~\ref{tab:eval-pfr} shows the normalized results. Manual mode reproduces the Blueprint graph with an F1 of 0.998: out of 403 Blueprint edges, only 1 is missed and 1 extra edge is produced. Edge style accuracy is 100\%. This confirms that KnowTeX's processing of \verb|\uses{}| and \verb|\proves{}| commands is fully compatible with the Lean Blueprint system.

Infer mode achieves an F1 of 0.826 without any annotations. The main limitation is recall (0.764): 95 Blueprint edges are missed because the PFR source does not use \verb|\ref{}| commands for all dependencies. Many connections in this text are expressed through mathematical notation and natural language rather than explicit cross-references. Of the 35 extra edges produced by infer mode, 8 are reachable through alternative paths in the Blueprint graph, bringing the path-based F1 to 0.864.

A notable difference appears in edge style classification. The Blueprint marks 62 edges as dashed (statement-level), while infer mode produces only 5 dashed edges. This gap arises because infer mode discovers most dependencies through \verb|\ref{}| commands inside proofs (rule D1, producing solid edges), whereas the Blueprint assigns statement-level status based on information from the Lean side that is not visible in the \LaTeX\ source.

\subsection{Discussion}

The evaluation shows that infer mode performs well on structured mathematical texts: it achieves high node detection accuracy and discovers the majority of true dependencies. The path-based analysis is particularly encouraging, as it shows that most extra inferred edges are transitively justified shortcuts rather than genuinely incorrect claims.

It is worth noting that the quality of the inferred graph is closely tied to the quality of the source text itself. The deterministic rules (D1--D2) rely on cross-references: an author who consistently uses \verb|\ref{}| when citing earlier results will produce a richer and more accurate dependency graph than one who refers to results only by name in running text. Similarly, the defined-term matching rule (D4) depends on authors marking new terms with \verb|\emph{}| or \verb|\textbf{}| inside definition environments, and the heuristic rules (H2--H3) assume that corollaries follow their parent theorems and that lemmas precede the results they support. When these conventions are followed, the inference rules capture dependencies effectively. Even assigning \verb|\label{}| to environments that are never cross-referenced improves the output, since it gives nodes meaningful names in the graph rather than numeric fallbacks. In short, the cleaner and more explicit the source text, the better the resulting dependency graph. KnowTeX thus rewards good mathematical writing practices, and its output can serve as indirect feedback on the structural clarity of a document.

The main weaknesses are: (a) intermediate structural concepts may be skipped when they are not explicitly referenced, (b) label naming inconsistencies inflate raw error rates, and (c) some edge types are misclassified. All of these can be corrected through the interactive edge review GUI, which allows users to inspect and fix inference results before exporting the final graph.

%% ============================================================
\section{Discussion and Broader Applications}
\label{sec:discussion}
%% ============================================================

While our experiments focus on mathematical research texts, most of the so-called hard sciences could benefit from a tool like KnowTeX. \LaTeX\ is used by physicists, chemists, biologists, and others, and the logical dependency structures that KnowTeX makes explicit could be beneficial to these disciplines as well.

As demonstrated in Case Study~3 (Section~\ref{subsec:eval-corpus}), KnowTeX can process corpora at a scale of hundreds of definitions and produce dependency graphs aligned with Wikidata identifiers. This combination of structural dependency data and semantic identifiers creates a resource that goes beyond visualization: it provides machine-readable metadata that could be consumed by downstream systems. Potential applications include prerequisite prediction for intelligent tutoring systems, concept similarity measures derived from graph distance, and training data for models that learn to infer mathematical dependencies from text. We have not pursued these applications in this work, but the benchmark infrastructure and evaluation data are included in the repository to support future research in this direction.

In practice, the generated graph serves as a feedback mechanism for the author. Reviewing the graph before finalizing a manuscript can reveal structural issues that are hard to spot in linear text: an important definition that no later result depends on, a theorem whose proof references only a small subset of the stated prerequisites, or a cluster of results that is disconnected from the rest of the chapter. These observations are not limited to inference errors; they often reflect genuine structural properties of the text that the author may want to address.

A key design decision in KnowTeX is to keep manual and infer modes complementary. Authors can start with infer mode to get an initial dependency graph, review and correct it through the GUI, and then export the corrected \verb|\uses{}| annotations back into their \LaTeX\ source. This workflow lowers the barrier to adopting dependency annotations: instead of annotating from scratch, authors only need to verify and fix automatically generated suggestions.

%% ============================================================
\section{Future Work}
\label{sec:future}
%% ============================================================

\paragraph{AI-assisted command generation.}
Inspired by Trouver, one possible enhancement would be to integrate machine-learning-based classification of mathematical environments (definitions vs.\ theorems vs.\ explanations) to improve the accuracy of the D4 and H4 rules. Such a feature could reduce manual corrections while preserving alignment with the original text structure.

\paragraph{LLM-assisted edge filtering.}
The error analysis in Sections~\ref{subsec:eval-corpus} and~\ref{subsec:eval-pfr} shows that a portion of the false positive edges are semantically incorrect matches caused by stem overlap or overly broad term matching. A natural extension would be to use a large language model as a post-processing filter: given a candidate edge and the text of the source and target environments, the model could judge whether the dependency is mathematically meaningful. This approach could reduce false positives without changing the rule-based core of KnowTeX, preserving its deterministic and reproducible nature while adding an optional refinement step.

\paragraph{Toward structured mathematical knowledge resources.}
The corpus-scale experiment in Section~\ref{subsec:eval-corpus} shows that KnowTeX can produce dependency-annotated data aligned with Wikidata identifiers. This combination of structural and semantic information could serve as a building block for constructing large-scale mathematical knowledge resources that do not yet exist in widely available form. For example, dependency edges could be enriched with ontological relations (such as ``is a special case of'' or ``generalizes'') to create structured datasets for prerequisite prediction or retrieval-augmented generation in mathematical question answering.

\paragraph{Improving inference quality.}
The evaluation reveals specific patterns where inference falls short, particularly for intermediate structural concepts and generic defined terms. Developing more sophisticated matching strategies, such as incorporating type information from the mathematical environment or using contextual embeddings, could address these limitations.

\paragraph{Rule configuration and multi-target proofs.}
Users may want to turn individual rules on or off, or change their precedence, to see how each rule contributes to the final graph; we plan to expose this as an option. Rule D3 currently reads a single reference in a proof header, so a proof of several results at once (``Proof of Theorems 2.4 and 2.5'') is linked to only one of them; extending D3 to link the proof to every target named in the header is straightforward and planned. Reporting how often each rule fires, and what it contributes, on each case study would also make the evaluation more transparent.

\paragraph{Direct input of text fragments and other formats.}
We are working toward a mode where short pieces of text can be pasted into a window and processed directly, which would extend the benefit of KnowTeX to anyone reading or studying a text, not only to authors of a full document. In the same direction, processing corpora in their original formats, including markdown, would allow experiments such as Case Study~3 to be reproduced with no \LaTeX\ conversion at all.

%% ============================================================
\section{Conclusion}
\label{sec:conclusion}
%% ============================================================

We have presented KnowTeX, a standalone \LaTeX-oriented tool for generating dependency graphs from mathematical texts. Compared to the first version of this preprint, which supported only manual annotations, the current system introduces an infer mode with eight layered rules (D1--D4 deterministic, H1--H4 heuristic) that automatically discover dependencies from cross-references, defined-term matching, and structural patterns.

Our evaluation on four case studies shows that infer mode achieves good accuracy on individual documents (F1 of 0.818 on group theory, 0.730 normalized on category theory, 0.826 on a Lean Blueprint project) with very few truly spurious edges. On a corpus of 611 definitions from the MathGloss dataset, the combined rules achieve an F1 of 0.885 with perfect recall, and an error analysis reveals that 254 of the 433 edges flagged as false positives (58.7\%) are genuine dependencies not annotated in the ground truth. Manual mode reproduces an existing Lean Blueprint dependency graph with an F1 of 0.998, confirming full compatibility with the Blueprint annotation system.

Dependency graphs themselves are not new; the contribution of this work is the way the graph is obtained. Infer mode extracts a dependency graph from ordinary \LaTeX\ with no annotation, no proof assistant, and no formalization. Existing tools either require manual annotation (the first version of KnowTeX, Lean Blueprint, plasTeXdepgraph) or work from inside a formal ecosystem (LeanArchitect, from the Lean side). KnowTeX thus produces a graph for an unannotated \LaTeX\ document where none was available before, and it stays compatible with Lean Blueprint so that this graph can feed into the existing formalization workflow rather than compete with it. Working from plain \LaTeX, with nothing required from the author, is what makes this entry point low-cost and usable by authors who do not work in a proof assistant at all. The tool is publicly available under the Apache~2.0 license.

%% ============================================================
%% Acknowledgments / Disclosure
%% ============================================================

%\paragraph{Acknowledgments.}
%% TODO: Add funding acknowledgments, if any.

\paragraph{Disclosure of Interests.}
The authors have no competing interests to declare that are relevant to the content of this article.

%% ============================================================
%% Bibliography
%% ============================================================

\bibliographystyle{alpha}
\bibliography{references}

\end{document}